# X-ray scattering investigation of hydride surface segregation in epitaxial Nb films

David A. Garcia-Wetten[1], Philip J. Ryan[2], Jong Woo Kim[2], Dominic P. Goronzy[1], Roger J. Reinertsen[1], Mark C. Hersam[1,4], Michael J. Bedzyk[1,5].

1. Department of Materials Science and Engineering, Northwestern University, Evanston, Illinois 60208, United States
2. Advanced Photon Source, Argonne National Laboratory, Argonne, Illinois 60439, United States
3. Department of Physics and Astronomy, Northwestern University, Evanston, Illinois 60208, United States
4. Department of Chemistry, and Department of Electrical and Computer Engineering, Northwestern University,  Evanston, Illinois 60208, United States
5. Department of Physics and Astronomy, Northwestern University, Evanston, Illinois 60208, United States

**Abstract**

Hydride precipitation in niobium-based, superconducting circuits is a damaging side-effect of hydrofluoric acid treatments used to clean and thin the Nb surface oxides and Si oxides. The precipitate microstructure is difficult to probe because of the high hydrogen mobility in the niobium matrix. In particular, destructive techniques used to prepare samples for elemental depth profiling can change the hydride structure. Here, we use X-ray surface scattering to non-destructively probe the depth distribution of precipitates in hydrided, epitaxial, niobium thin films. We find that the niobium hydride is confined within the top 10 nm of the surface, and that the domains are discretely tilted with respect to the film's surface crystallographic orientation.

## I.    Introduction

Superconducting circuits are a promising physical implementation of quantum computing. While superconducting qubit coherence times have dramatically improved due to advancements in device architecture, materials defects arising during fabrication are still a notable source of decoherence.[1, 2] Niobium thin films are a common material for superconducting qubits due to their ease of deposition and relatively high superconducting transition temperature. However, this layer necessarily introduces a lossy, surface niobium oxide layer, and current state-of-the-art qubits employ various surface treatments to reduce or remove this oxide. Notably, hydrofluoric acid treatments are commonly used with much success [3] to thin this oxide layer. Fluorine-based chemical etchants are now routinely used as the last step before measuring resonators [4]. Note that Nb is known as an excellent hydrogen getter and has even been considered a hydrogen storage material [5]. This acid etch process introduces a significant amount of hydrogen into the Nb film, which is an origin of quasiparticle losses in superconducting circuits [6, 7].

Superconducting qubits are operated at millikelvin temperatures. When these circuits are cooled, the solubility of H in the Nb matrix drops to nearly zero, causing the formation of highly concentrated, highly strained hydride precipitate phases. Even without an HF treatment step, the predominance of $H_2$ in the base pressure of most vacuum systems introduces non-negligible concentrations of H into Nb films. In bulk Nb, these precipitates are incoherent and cause the formation of large dislocation networks [8]. However, the structure of these precipitates in Nb thin films is unclear. Here, we point out the conflicting picture of the NbH precipitate structure in thin films in literature. The most prevalent hypothesis is the formation of columnar NbH grains, supported by finite element analysis [5], and transmission electron microscopy (TEM). However, these reports conflict with depth profiling measurements, such as time-of-flight secondary mass spectrometry (ToF-SIMS) [9], which suggest that the NbH precipitates are found at the film surface. Phase-field simulations of the initial stages of NbH precipitation in a Nb matrix near a free surface have suggested that precipitates formed in the bulk quickly migrate towards the surface[10].

We will show that this discrepancy of surface segregation versus columnar formation is due to the destructive nature of these two measurements (TEM and ToF-SIMS). We employ the non-destructive nature of X-ray scattering for a multi-length-scale investigation of the formation of NbH precipitates at low temperatures. Grazing incidence X-ray scattering is used to produce a depth-penetration profile of the NbH precipitates. We find that the NbH precipitates appear within 10 nm of the surface, confirming the predictions of phase-field simulations that these structures find the free surface as they form. Detailed mapping of the NbH reflections in reciprocal space is also used to identify the formation of low-angle grain boundaries between the precipitates. This work aims to 1) provide an accurate model of the NbH precipitate structure to better associate the precipitate occurrence and superconducting circuit performance and 2) elucidate the nature of hydrogen precipitation in thin film metal systems. While our experiments are focused on thin films, our findings are also relevant to bulk Nb crystals.

## II. Experimental Methods
### a. Thin film growth and Synchrotron X-ray diffraction

50 nm thin films of Nb (110) were fabricated via molecular beam epitaxy (MBE) on a-plane α-$Al_2O_3$ (110). The 10x10x0.5 $mm^3$ sapphire substrate from MTI was degreased and annealed in air at 1100°C for 2 hours to form atomically smooth terraces on the surface of the substrate. It was degassed in the MBE UHV chamber (base-pressure = $2x10^{-10}$ Torr) at 200°C for 2 hours. The Nb film was deposited at a rate of 0.5 nm/min and a substrate temperature of 750°C. A 6 mm diameter, 99.99% pure Nb rod from American Elements was used in a Focus GmbH EFM 3 e-beam evaporator. The film was kept in UHV, annealed at 1000°C for 30 minutes, and then cooled to RT. All temperature ramp rates were 10°C/min.

The Nb-coated sapphire substrate was cleaved into smaller 5x5 $mm^2$ pieces for hydrogen incorporation. Acid etches were used to introduce hydrogen into the Nb layer as described elsewhere [6]. One piece was kept as a control, a second piece was dipped into 6:1 $NH_4F$:HF buffered oxide etchant (BOE) for 6 minutes, and a third piece was dipped into 8% HF for 4 minutes. The BOE treatment served to mimic relevant qubit fabrication processes, such as BOE dips before packaging. The HF treatment served to produce a reference sample with all relevant phases present at room temperature [11]. Acid etches such as these are routinely used to remove surface contamination and oxidation on Nb superconducting qubits before cryogenic microwave measurements [3].

X-ray scattering data was collected at the Advanced Photon Source (APS) 6-ID-B beamline at an incident photon energy of 10.00 keV (wavelength λ = 1.240 Å), with the samples held at room temperature or 5 K. The samples were cooled using an Advanced Research Systems 4K displex mounted on a Ψ-6-circle Huber diffractometer. A Pilatus 100K area detector was used to capture 3D reciprocal space maps (RSM).

Specular rod scans were performed using the typical θ−2θ scan motion. Off-specular rod scans were performed in constant incident angle mode. This grazing incidence XRD (GIXRD) was performed by fixing the χ circle at 90°, the μ circle at 0°, and constraining the incidence angle α. A coupled motion of the ν, φ, δ circles kept the scan direction confined to selected crystal truncation rods (CTR) for the Nb (110) surface. [12]

## III. Results and Discussion
### a. X-ray Scattering

Specular X-ray scattering data of the control, BOE, and HF-dipped Nb films presented in Figure 1(A,C,E) show line-scans of the second-order (220) substrate and film peaks at room temperature (RT) and 5 K. These specular θ−2θ scans are plotted as a function of the scattering vector magnitude Q = 4πSinθ/λ. The Nb reflection around 5.35 $Å^{-1}$ is the α-NbH phase, which consists of a BCC Nb lattice with a low concentration of H occupying tetrahedral interstitials. The very weak Bragg peak at Q = 5.13 Å$^{-1}$ corresponds to a Nb hydride β (room temperature) or ε (5 K) phase. These phases consist of a face-centered orthorhombic Nb lattice with ordered H at tetrahedral interstitial sites. [11] Because H is such a weak scatterer, XRD is not sensitive in structure factor to the rearrangement of H within the distorted, orthorhombic unit cell and cannot be used to distinguish the β phase from the ε phase. Both phases appear under X-ray scattering as a strained BCC Nb lattice. (Throughout this report, we will only use BCC *hkl* indices for Nb and its hydride phases.) Increased loading of H in the Nb α phase due to the HF treatments causes the Nb lattice to expand, as seen in the shift of the Nb (220) peak to lower Q. At room temperature, the BOE treated sample contains α phase Nb hydride. The HF-treated sample contains both the α and β Nb hydride phases. After cooling to 5 K, the scattering pattern changes due to H segregation. Based on the Nb-H bulk crystal equilibrium phase diagram [13], the solubility of H in the α-Nb phase is extremely low at 5 K.

The Al$_2$O$_3$ (110) peak shows a very small shift due to thermal contraction. The much larger shift in the (220) peak for the BCC NbH α phase upon cooling is primarily due to the rearrangement of H within the two-phase film and a smaller contribution due to thermal contraction. The broad Bragg peak for the low-temperature NbH ε phase occurs at the same Q as the room-temperature NbH β phase. Figures 1(B,D,F) show 2D cuts of the reciprocal space map of the second-order substrate and film reflections at 5 Kelvin sliced with the in-plane $Q_x$ direction along the [-112] direction. The diffuse ε phase at low $Q_z \sim 5.1$ Å$^{-1}$ is more prominent in this perspective. The ε phase reflection from the control and BOE-treated samples at 5 K is quite weak in intensity due to its diffuse and broad extent. The appearance of H in the untreated film is consistent with Lee et al. [9] and Torres and Goronzy et al. [6]. As more hydrogen is loaded into the sample, this peak becomes more intense and broader in $Q_x$. The vertical streaks in Figures 1(B,D,F) about the Nb (220) reflection in each sample have been previously observed by X-ray scattering [14, 15], but not with such clarity and to such high order. They are periodic in $Q_x$ and are proposed to be due to a periodic strain field in the Nb thin film caused by the lattice mismatch at the interface with the sapphire substrate. The real-space periodicity corresponds to 16 nm. This feature is a surprising result, and while we hope to study it in future work, we do not believe they are related to the formation of hydride structures we observe in this study.

A lateral slice in reciprocal space of the ε-NbH reflection of the HF-treated film is shown in Fig. 2. The white arrow is along the $Q_x$ or [-112] direction plotted in Figures 1(B,D,F). This cut reveals the anisotropy of the ε-NbH reflection within the plane. Surrounding a central peak are two lobes in the [-110] direction. The broadening of the main reflection and the two lobes was found purely due to mosaicity, with negligible broadening due to a size effect or diffuse scattering. Figure 3 shows evidence for this mosaicity by comparing the second-order (220) reflection with the first-order (110) reflection and off-specular (1-12) reflections. The broadening occurs along the Ewald sphere and is constant in angle as opposed to constant in Q. Mosaicity of the side lobes indicates that there is a discrete tilting of the concentrated NbH phase by ±1.1°, which has been observed before in TEM of NbH foils and is consistent with tilting towards the <110> direction [8]. The existence of low-angle grain boundaries between precipitates within the single-crystalline Nb matrix suggests that the NbH domains are tilted by the pile-up of dislocations.

### b. Surface Segregation

To demonstrate that destructive profiling modifies the NbH precipitate structure, we have performed ToF-SIMS of the HF-treated film with primary Cs+ and O- ions. Figure 4 shows the measured Nb, Al$_x$O$_y$, and Nb$_x$H$_y$ containing species as a function of sputtered depth. The depth scale is calibrated from the thickness as measured by XRD to the onset of the Nb/Al$_2$O$_3$ interface. As measured with Cs$^+$ primary ions, the hydrogen content is observed within the top 15 nm of the film. However, when measured with O$^-$ primary ions, the H content is observed mostly within the bulk of the film down to 40 nm. This effect has been observed in the TaH system by Asakawa et al.[16] Movement through and out of the metal phase is induced by ion impingement and the removal of the NbO$_x$ surface oxide, which acts as a H barrier in this system. This H diffusion hinders an accurate measurement of the depth profile.

Due to a less-than-one index of refraction, X-rays undergo total external reflection. At low incident angles, the penetration depth of X-rays can be dramatically reduced to a couple of nanometers due to the evanescent wave effect [17, 18]. One can take advantage of this fact to selectively probe the surface of a material by restricting the interaction volume of the incident X-ray beam to only include the near-surface region. The X-ray penetration depth is given by:

$$\Lambda = \frac{\lambda}{4\pi}\left(Re\left\{[\alpha_c^2 - \alpha^2 + 2i\beta]^{\frac{1}{2}}\right\}\right)^{-1}$$

Where $\lambda$ is the incident X-ray wavelength, $\alpha_c$ is the material critical angle, $\alpha$ is the incidence angle, and $\beta$ is the complex component to the index of refraction. For Nb, $\alpha_c = 0.32°$ at 10.00 keV and $\Lambda_{min} = 1.2$ nm.

GIXRD measurements of the BOE and HF-treated samples were made at fixed incidence angles about the Nb critical angle to make a series of measurements of the phase ratio of ε-NbH to α-Nb as a function of depth below the surface of the film. Figure 5 shows this ratio plotted against X-ray penetration depth for the BOE and HF-treated samples. The ratio was calculated by summing the integrated intensity of the off-specular ε-NbH and α-Nb (121) reflections. The signal for the BOE-treated film is multiplied by 20 to show the curves for both the BOE and HF-treated samples on the same scale. The inset shows an example of the 3D RSM plotted with boxes outlining the extent to which the total intensity was summed. The Nb peak diminishes rapidly as the penetration depth is decreased. This strongly suggests that the NbH concentrated phase at 5 Kelvin lies within 10 nm of the free surface of the film as it precipitates. The fact that the line shapes of the profiles of the BOE and HF-treated samples are identical suggests that as more hydrogen is introduced, the precipitates do not extend deeper into the film but become more populated at the surface.

IV. Conclusion

We have non-destructively investigated the low-temperature structure of niobium hydride precipitates in single-crystal, niobium thin films. Motivated by contrary results in the literature, we demonstrated that destructive techniques provide conflicting answers as to the location of NbH precipitates within a Nb thin film. In agreement with phase-field simulations, which predict the initial precipitation predominantly at the film's free surface, we determined that the NbH precipitates are found within 10 nm of the free film surface using GIXRD. Additionally, we identified evidence of dislocation pile-up between each precipitate from the formation of discrete domain tilts. We must note again that this domain structure formed in a single-crystal Nb matrix. Many Nb films used for superconducting circuits in the literature are polycrystalline or textured, and the defects present may cause the precipitate domain shapes to be different than the ones in the films studied here. The formation of a highly disordered normal metal at the surfaces of superconducting circuits is likely the culprit of quasiparticle loss in hydrogen-loaded devices [6, 7]. This power-independent, quasiparticle-related loss has also been confirmed in hydride tantalum-based circuits [19]. Tantalum is a similar superconductor with a nearly identical crystal structure and a similar affinity to uptake hydrogen. Furthermore, normal metal at the surface of superconducting resonators decreases the effective superconducting gap and increases power-independent losses [20]. However, the evidence for dislocation pile-ups due to NbH precipitation also warrants a future investigation into the role dislocations play as a loss source in general.

V. Acknowledgments

We thank Peter Voorhees and Tyler Leibengood for discussions on niobium hydride phase formation. This work was supported by the U.S. Department of Energy, Office of Science, National Quantum Information Science Research Centers, Superconducting Quantum Materials and Systems Center (SQMS) under Contract No. DE-AC02-07CH11359. This research used resources of the Advanced Photon Source, a U.S. Department of Energy (DOE) Office of Science user facility operated for the DOE Office of Science by Argonne National Laboratory under Contract No. DE-AC02-06CH11357. This work made use of the

Jerome B.Cohen X-Ray Diffraction Facility supported by the MRSEC program of the National Science Foundation (DMR-2308691) at the Materials Research Center of Northwestern University and the Soft and Hybrid Nanotechnology Experimental (SHyNE) Resource (NSF ECCS-2025633.)

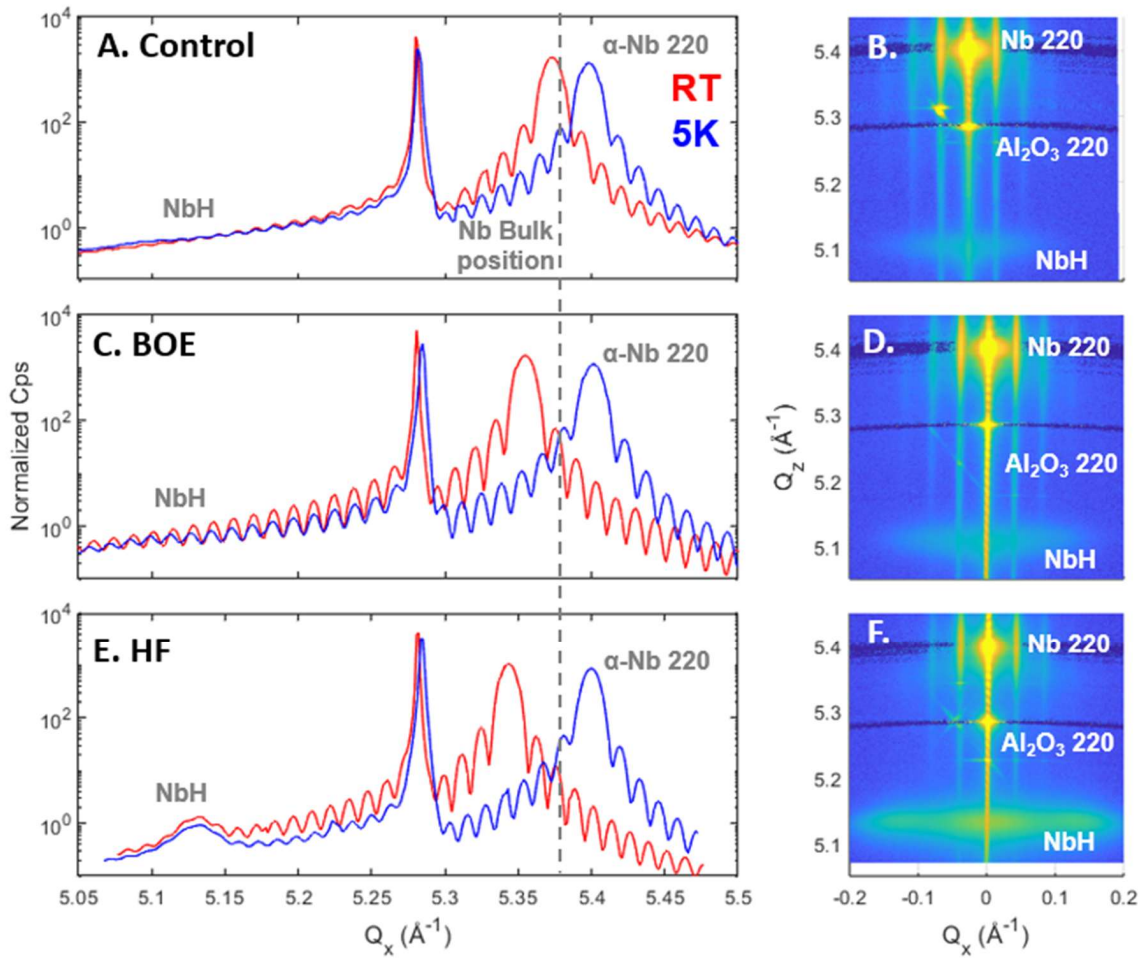

**Figure 1.** X-ray scattering patterns of the second-order specular reflections: A., C., and E. show line-scans, respectively of the control, BOE-treated, and HF-treated films at room temperature and 5 Kelvin. The Nb bulk lattice parameter is shown with a dashed line. B., D., and F. show reciprocal space maps of these films, respectively, taken at 5 Kelvin.

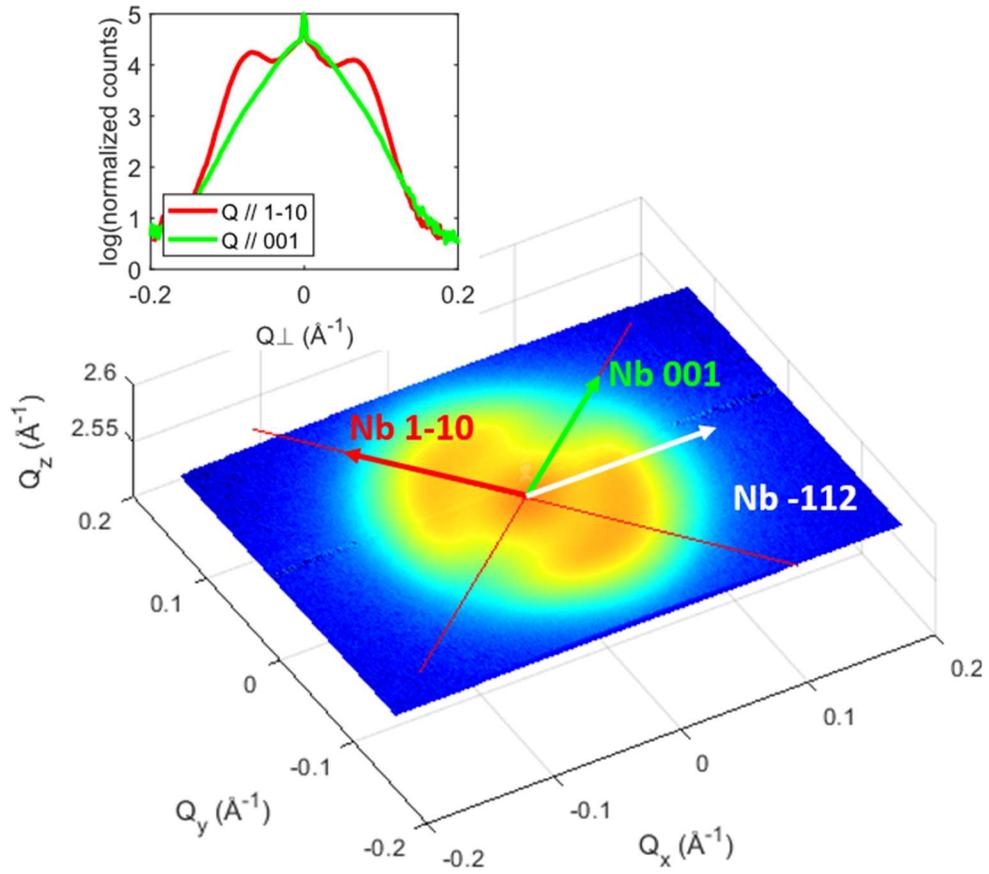

**Figure 2.** Lateral slice of NbH concentrated phase, first order specular reflection. The reflection is asymmetric in two ways: the central peak is narrower in the [001] direction and has lobes in the [1-10] direction. The inset shows the intensity profiles in these two perpendicular directions

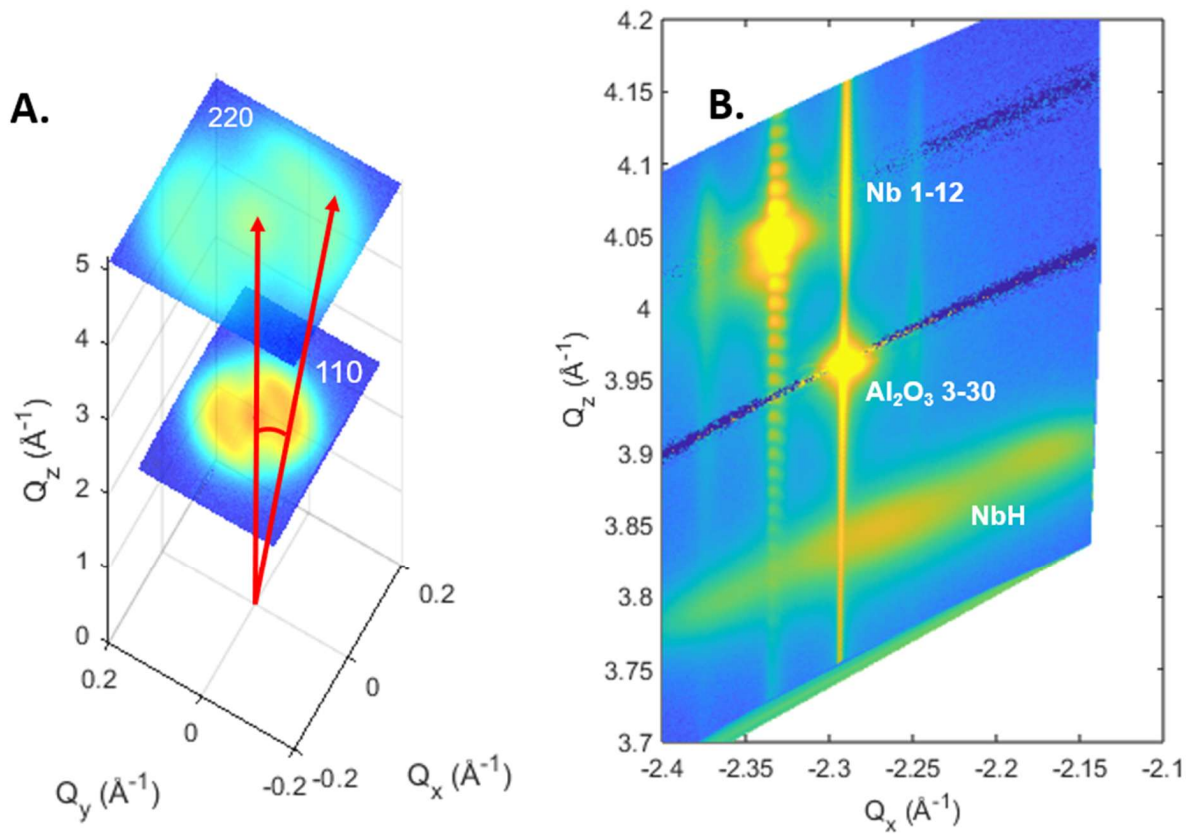

**Figure 3.** Mosaic spread of the NbH concentrated phase: A. The first- and second-order reflections have constant width in the rocking ω angle, not Q. Red arrows indicate this constant angle from the (000) point. B. Reciprocal space map of Nb and $Al_2O_3$ off-specular reflections. The broadening of the NbH reflection is along the Ewald sphere.

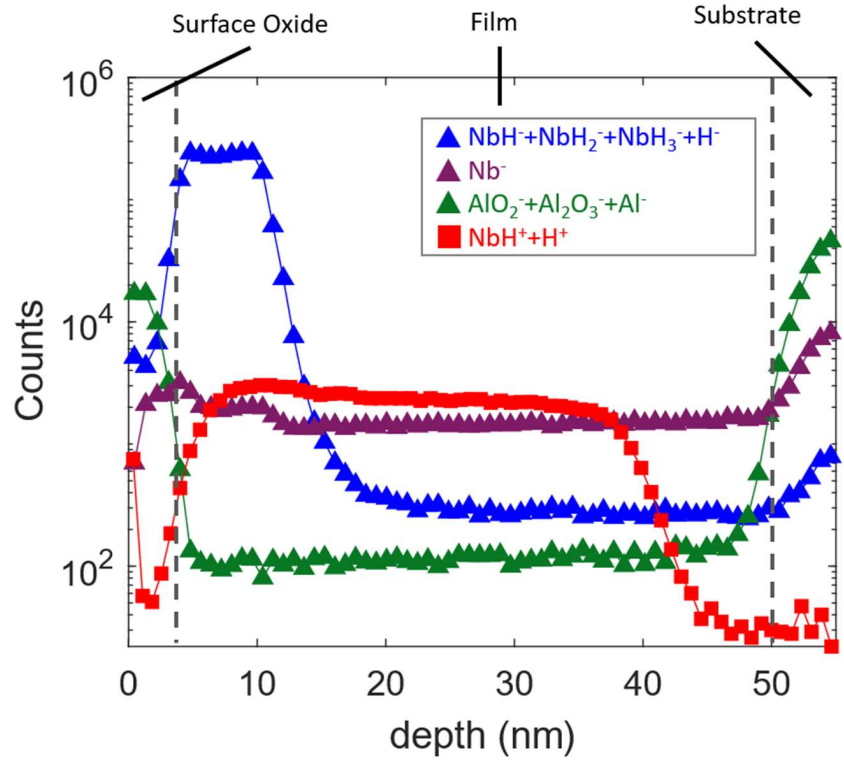

**Figure 4.** ToF-SIMS of the HF-treated film. The triangle-labeled profiles were measured with the use of Cs⁺ primary ions, and the square-labeled profile was measured with the use of O⁻ primary ions.

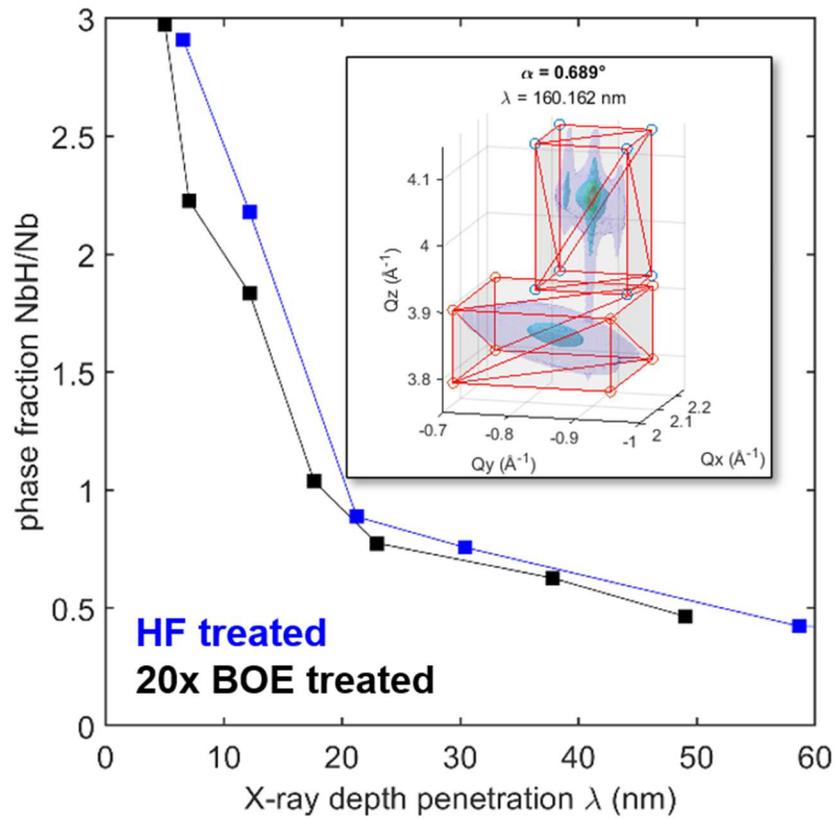

**Figure 5.** Depth profile of the NbH phase fraction from GIXRD. The HF-treated film is shown in blue, and the BOE-treated film is shown in black x20 to be more easily compared. Each data point is the ratio of integrated intensities of the NbH and Nb reflection from a single rod scan across the Nb 211 reflection. The inset shows an example of the scan regions used for these calculations.